\documentclass[sigconf]{acmart}

\usepackage{multirow}
\usepackage{amsmath}

\usepackage{tcolorbox}
\usepackage{xcolor}
\usepackage{caption}  
\usepackage{subcaption}

\definecolor{promptcolor}{RGB}{233,233,233}

\newcommand{\llmprompt}[1]{%
  \begin{tcolorbox}[colback=promptcolor, colframe=promptcolor, boxrule=0.5pt, arc=3pt, left=8pt, right=8pt, width=1.0\linewidth]
    {\ttfamily\small #1}
  \end{tcolorbox}%
  \captionsetup{type=figure, justification=raggedright}
}

\AtBeginDocument{%
  }

\setcopyright{acmcopyright}
\copyrightyear{2018}
\acmYear{2018}
\acmDOI{XXXXXXX.XXXXXXX}

\acmConference[Conference acronym 'XX]{Make sure to enter the correct
  conference title from your rights confirmation emai}{June 03--05,
  2018}{Woodstock, NY}
\acmPrice{15.00}
\acmISBN{978-1-4503-XXXX-X/18/06}

\begin{document}

\title{Semantic Ranking for Automated Adversarial Technique Annotation in Security Text}


\author{Udesh Kumarasinghe}
\affiliation{%
  \institution{Purdue University, IN}
  \country{USA}
}

\author{Ahmed Lekssays}
\affiliation{
  \institution{Qatar Computing Research Institute, HBKU}
  \country{Qatar}
}

\author{Husrev Taha Sencar}
\affiliation{
  \institution{Qatar Computing Research Institute, HBKU}
  \country{Qatar}
}

\author{Sabri Boughorbel}
\affiliation{
  \institution{Qatar Computing Research Institute, HBKU}
  \country{Qatar}
}

\author{Charitha Elvitigala}
\affiliation{
  \institution{University of Colombo}
  \country{Sri Lanka}
}

\author{Preslav Nakov}
\affiliation{
  \institution{MBZUAI}
  \country{UAE}
}


\begin{abstract}
We introduce a novel approach for mapping attack behaviors described in threat analysis reports to entries in an adversarial techniques knowledge base.  Our method leverages a multi-stage ranking architecture to efficiently rank the most related techniques based on their semantic relevance to the input text. Each ranker in our pipeline uses a distinct design for text representation. To enhance relevance modeling, we leverage pretrained language models, which we fine-tune for the technique annotation task. While generic large language models are not yet capable of fully addressing this challenge, we obtain very promising results. 
We achieve a recall rate improvement of +35\% compared to the previous state-of-the-art results.
We further create new public benchmark datasets for training and validating methods in this domain, which we release to the research community aiming to promote future research in this important direction.

\end{abstract}




\maketitle

\section{Introduction}

Understanding threat behaviors is essential for developing effective cyber defense capabilities. 
Acquiring this crucial knowledge involves analyzing a vast array of artifacts in the aftermath of security incidents. 
Through such analyses, experts reconstruct the sequence of events that occurred during the incident, uncover the tactics, techniques, and procedures employed by threat actors, and determine the extent and impact of the attacks on data and systems.
The insights and findings of these experts are typically documented in the form of unstructured or semi-structured natural language text and shared with the broader security community \cite{symantec_2023,Securelist,Broadcom2023,WithSecure2023}. Other analysts, with access to resulting threat intelligence through public or paid platforms, evaluate the information to strengthen existing detection, mitigation, and remediation capabilities.

The acquisition of threat knowledge from security text has predominantly relied on highly expert-driven and time-intensive processes. 
Recognizing the importance of this challenge to security operations, several approaches have been proposed to automatically extract threat behaviors from threat intelligence reports.
These efforts have focused on different objectives, such as  extracting indicators of compromise (IoCs) \cite{liao2016acing, zhu2018chainsmith}, characterizing adversary tactics and techniques \cite{ttpdrill:acsac:2017,legoy2020automated, tsai2020cti, you2022tim, ladder:alam:2022, attackg:esorics:2022}, identifying threat actions \cite{husari2017ttpdrill, tsai2020cti, gao2021enabling, alam2022looking}, and creating cybersecurity knowledge graphs \cite{gao2021system,gao2021enabling,ji2022knowledge,ren2022cskg4apt}.

In this work, our focus lies in annotating threat behaviors described in a threat intelligence report with reference to \textcolor{black} {adversarial techniques described in the} MITRE's ATT\&CK knowledge base \cite{mitre-attack}. 
This framework serves as a widely adopted resource for analyzing and understanding the tactics, techniques, and procedures (TTPs) deployed by adversarial threat actors. 
It provides a comprehensive and standardized way of describing various stages of an attack, from initial access and reconnaissance to execution and post-exploitation. 
By aligning the findings of security reports with pertinent tactics and techniques, security analysts and researchers can promptly identify patterns and trends in  in the way attacks are carried out. 
Furthermore, as adversaries cannot change their techniques and procedures rapidly,  these associations aid in attributing attacks to specific threat actors and in clustering reports into campaigns.
Addressing these requirements, several security vendors and organizations that generate and assess threat intelligence reports also identify the adversarial techniques employed during investigated attacks as part of their insights\footnote{These encompass threat intelligence entities like WeLiveSecurity, TrendMicro, AT\&T Security Labs, MITRE, and CISA. A more comprehensive discussion of this data is presented in Section \ref{sec:dataset}.}.

Previously proposed solutions for the technique annotation task have so far yielded limited practical impact, primarily due to two key factors. Firstly, some of these approaches framed the task as a classification problem. A significant limitation of this formulation lies in the lack of extensive ground truth datasets crucial for effectively training multi-class classifiers capable of distinguishing the correct technique among hundreds of distinct adversarial techniques. Additionally, classification models tend not to excel in scenarios involving human-in-the-loop, where the top-k techniques are presented to the annotator for selection.
The second factor pertains to the constraint of conventional natural language processing (NLP) methods in adeptly capturing the contextual nuances of mentioned entities and their interrelationships within the text.
The prevalence of domain-specific terms, complex sentence structures \cite{satvat2021extractor}, improper punctuation \cite{mu2018understanding}, and the amalgamation of diverse data formats (including code, tables, figures, and non-standard textual elements like IP/MAC addresses, domain names, and file hashes) collectively undermines the interpretation of reports and the extraction of attack behaviors.
Moreover, the absence of standard benchmark datasets led to the use of closed, custom datasets, thereby making it difficult to compare these approaches on an equal footing.

To address these issues holistically, this study introduces a new problem formulation that better aligns with the practical aspect of the problem.
In pursuit of this goal, we devise a multi-stage ranking pipeline \textcolor{black}{\cite{matveeva2006high,asadi2013effectiveness,clarke2016assessing,mackenzie2018query}} that assesses the semantic relevance between passages of text within a report and the corresponding reference text in the ATT\&CK knowledge base, that provides a data-driven understanding of adversary techniques.
In addition, we introduce a comprehensive public dataset including passages extracted from incident reports, along with manually annotated technique IDs\footnote{For anonymity, the link to the dataset is omitted.}. 

\subsection{MITRE ATT\&CK Framework}
The ATT\&CK framework consists of a structured matrix that categorizes and organizes the TTPs based on real-world observations\footnote{\textcolor{black}{\hyperlink{https://attack.mitre.org/versions/v12/matrices/enterprise/}{https://attack.mitre.org/versions/v12/matrices/enterprise/}}}.
It provides a comprehensive and systematic representation of various attack vectors and serves as a reference guide for understanding and mitigating potential threats. 
The matrix is divided into rows and columns, with columns representing 14 distinct tactics that represent high-level goals or objectives that adversaries aim to achieve, such as "Initial Access," "Execution," and "Exfiltration."
Each tactic is further populated with specific techniques that adversaries employ to execute those tactics, with the associated techniques listed in rows beneath each column.
Techniques offer detailed insights into how attackers maneuver through an organization's defenses. 
For further granularity, some techniques are further broken down into "subtechniques" that provide a deeper understanding of the specific variations and nuances of an approach. Additionally, "procedures" refer to the step-by-step sequences that adversaries follow to execute their techniques. 
This structured and standardized vocabulary equips cybersecurity professionals with a shared framework to effectively analyze, respond to, and defend against cyber threats.
As an evolving framework, ATT\&CK is currently in its 13th version, adapting to changes in the threat landscape.

The ATT\&CK framework encompasses a collection of 193 techniques and 401 subtechniques, each meticulously documented and categorized.
Each technique is assigned a unique ID and a descriptive title. Technique descriptions can also encompass potential detection and mitigation strategies, along with the required data sources for identifying such behaviors. 
Furthermore, some technique descriptions include concise summaries of sample procedures with reference to their source threat analysis reports. 
These procedures are often detailed accounts of actions carried out by threat actors. Our aim is to map these procedural descriptions to overarching technique descriptions.
Table \ref{table:mitre_stats} presents statistics on the number of reports used in the extraction of threat behaviors underpinning the knowledge base.

It's important to highlight that not all techniques hold equal likelihood of being mentioned in a report. Reports are essentially generated based on the analysis of attack artifacts. Techniques without traces, such as those within the "Reconnaissance" or "Resource Development" tactics, or those that evade detection due to the absence of necessary data sources, may not be highlighted. 
In contrast, techniques associated with "Initial Access" are more commonly found in these reports. This is because, during this stage, attackers have limited control over the compromised systems, making it difficult to conceal their traces. 


\begin{table}
\caption{ATT\&CK Matrix Statistics}
\label{table:mitre_stats}
\begin{tabular}{lccc}
    \toprule
    \bf Resource Type & \bf Top-level & \bf Sub-level & \bf Total \\
    \midrule
    Techniques & 193 & 401 & 594 \\
    Techniques w/o references & 38 & 71 & 109 \\
    Referenced reports & 700 & 607 & 1307 \\
    \bottomrule
\end{tabular}
\end{table}

\subsection{About Our Approach}

\begin{figure*}[ht]
\centering
\begin{subfigure}{0.49\linewidth}
    \includegraphics[width=\linewidth]{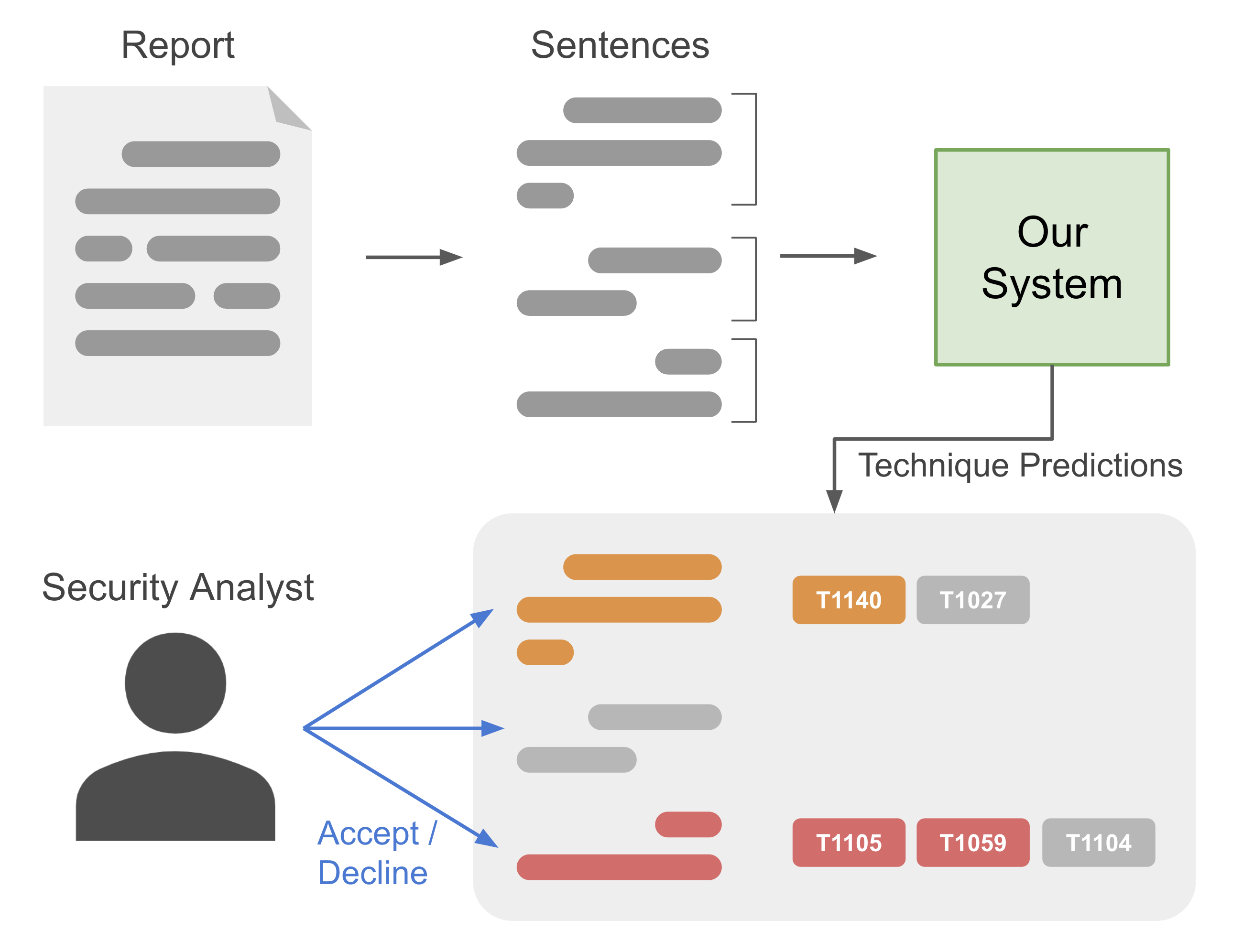}
\caption{}
\label{fig:usecase}
\end{subfigure}
\begin{subfigure}{0.49\linewidth}
\includegraphics[width=\linewidth]{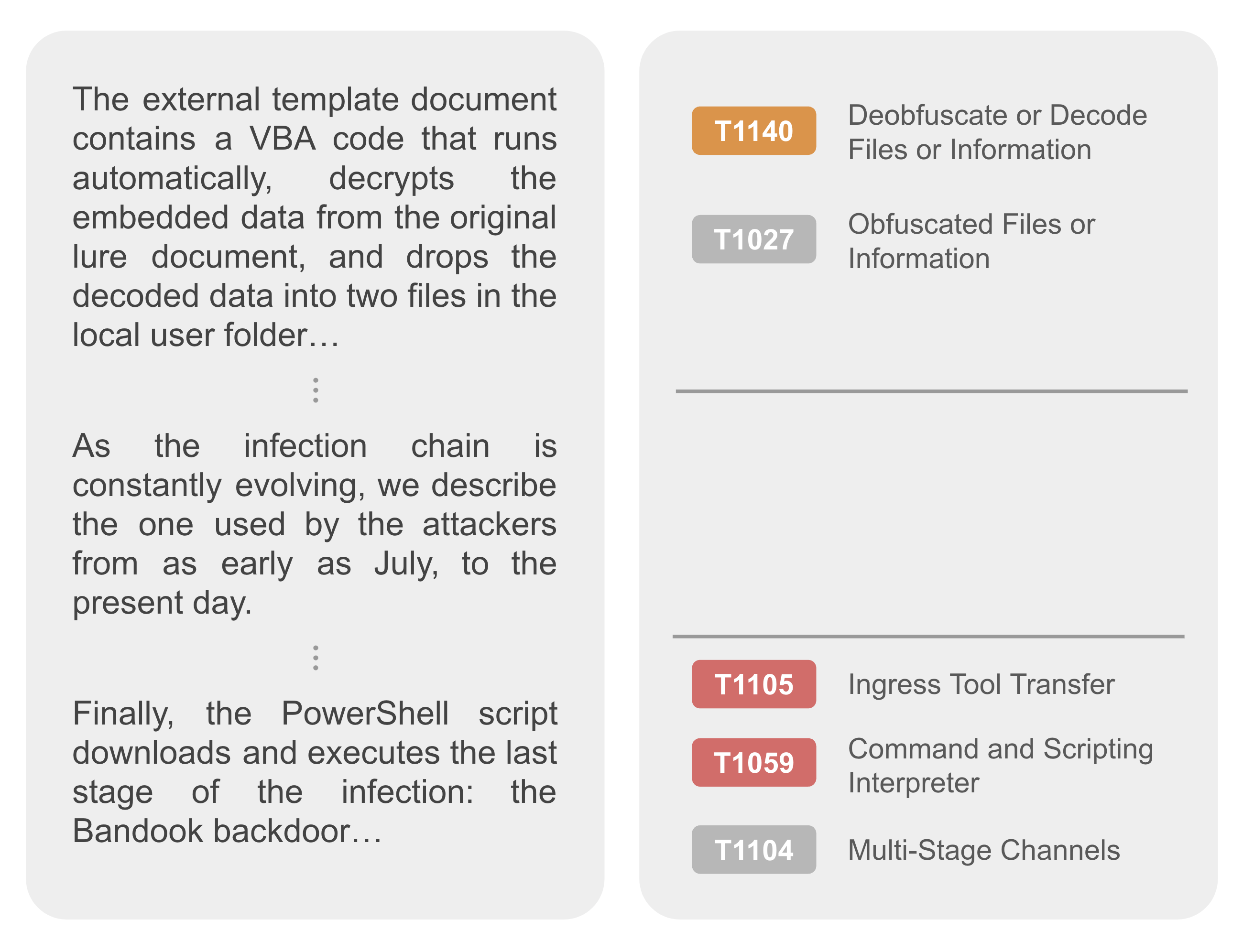}
\caption{}
\label{fig:examples}
\end{subfigure}
\caption{\textcolor{black}{(a) The practical application of our proposed system in annotating threat reports, (b) Examples of report sentences\protect\footnotemark ~with corresponding predicted techniques.}}
\label{fig:usecases_examples}
\end{figure*}

Our approach employs a multi-stage ranking architecture that strategically integrates diverse ranking models, optimizing the trade-off between effectiveness and efficiency. 
 In this multi-stage ranking framework, the candidates singled out by each stage become inputs for the subsequent ranking stages. Following the initial ranking, each stage receives a ranked list and refines it further by retaining only the most highly ranked candidates. 
The ranked list generated by the last stage of the multi-stage ranking architecture is the output of the system.
Our system takes a query text and a text corpus sourced from the ATT\&CK knowledge base as its input. 
\textcolor{black}{The use case for our system is depicted in Fig. \ref{fig:usecases_examples}.}

\footnotetext{\hyperlink{https://research.checkpoint.com/2020/bandook-signed-delivered}{https://research.checkpoint.com/2020/bandook-signed-delivered}} 

The design of our rankers leverages the capabilities of transformer-based pretrained language models to accurately evaluate the semantic relevance between the query and the text items in our corpus. \textcolor{black}{This approach aligns with methods used in web search, as described in \cite{zou2021pre, yan2021unified, lin2022pretrained,huang2020embedding}}. 
To optimize these ranking models and verify our findings, we utilized a newly generated dataset for fine-tuning and validation purposes.
Additionally, in our evaluation, we examined the performance of several large language models in the zero-shot learning scenario.
In summary, our contributions in this work include the following key aspects:

\begin{itemize}
\item We present a novel learning-to-rank approach designed for the annotation of threat behaviors outlined in 
threat intelligence reports, aiming to identify the most semantically relevant techniques within the ATT\&CK framework.

\item We constructed a new dataset of threat behaviors, enriched with human-annotated technique ID labels. 
This dataset is expected to facilitate the advancement of methods in this domain.

\item We provide a comprehensive benchmark that assesses the effectiveness of both open and closed large language models for the technique annotation task. 

\item We report new state-of-the-art performance results, indicating significant improvements over 
both previous technique annotation methods and the utilization of large language models for this task.
\end{itemize}


\section{Problem Formulation and System Overview}
\subsection{Formalizing Text Annotation}
Given a text excerpt extracted from a threat intelligence report, the goal of the text annotation task is to generate an ordered list of MITRE ATT\&CK Technique ID's, denoted as $T_i$'s, 
based on their relevance to the query text.
To achieve this, we adopt a text ranking formulation that leverages the semantic similarity between technique descriptions and the query.
Formally, we represent the set of technique IDs as $\mathcal{T}$ where $\mathcal{T}=\{T_1, T_2, \ldots , T_n\}$, and $n$ is the number of distinct adversarial techniques we aim to identify.
The descriptions of all techniques are contained in a large corpus $\mathcal{C}=\{it_1, it_2, \ldots , it_m\}$, where each $it_i$ is a text item associated with a technique $T$.
The association between the technique IDs and their descriptions is expressed through a function $f: \mathcal{T} \rightarrow \mathcal{P}(\mathcal{C})$, 
where $\mathcal{P}()$ denotes the power set, representing the set of all possible subsets of an input set.
In other words, the function $f$ maps each technique ID in the set $\mathcal{T}$ to a subset of text items in the set $\mathcal{C}$, i.e., $\forall T\in\mathcal{T}, f(T)\in\mathcal{C}$ . 
The text items associated with a particular technique, $f(T)$, consist of information collected from the ATT\&CK knowledge base about each technique. 
This information includes the technique's title, description, and the necessary measures for its detection and mitigation.

Formally, our task can be expressed as follows: given a corpus $\mathcal{C}$ and a query $q$, 
we aim to generate a ranked list of the top-$k$ technique IDs from $\mathcal{T}$ that maximizes a specific similarity metric, where $k$ represents the length of the ranked list.
The query $q$ may consist of a phrase or a passage, containing several sentences, extracted from a threat intelligence report.
The ranking model sorts the text items by calculating the probability of each text item entailing the query within its description.
To achieve this, we employ pretrained transformer-based models, leveraging the powerful language understanding capability of language models (LMs).
Specifically, our objective is to learn and rank dense representations of texts by estimating the probability:
\begin{equation}
P(related=1 | it_i, q) \triangleq \phi( \eta(it_i), \eta(q)) \text{ where }it_i\in\mathcal{C}.\\
\label{eq:Def1}
\end{equation}
This probability indicates the relevance of a text item with respect to a given query. 
In this formulation, $\eta()$ is a transformer-based encoder that maps a sequence of text tokens into a dense vector representation, potentially spanning hundreds of dimensions, with each dimension holding non-zero values. 
The function $\phi()$ is a symmetric measure used to evaluate the relevance between the encoded text items and the query.  
Alternatively, the probability in Eq. (\ref{eq:Def1}) can be estimated within an interaction-based model using the same learning architecture by jointly encoding the text and the query as:
\begin{equation}
P(related=1 | it_i, q) \triangleq \phi( \eta(it_i, q)) \text{ where }it_i\in\mathcal{C}.\\
\label{eq:Def2}
\end{equation}
In this context, $\phi$ can also be interpreted as a classifier, with its decision confidence serving as a ranking score.
It must be noted that, multiple text items are associated with each technique.
Therefore, the probability of a particular technique ID is determined by taking 
the maximum over all related text items in Eqs. (\ref{eq:Def1}) and (\ref{eq:Def2}) as follows:
\begin{equation}
P(T_i)=\max_{it} P(related=1 | it, q) \text{ such that } it\in f(T_i),\\
\label{eq:Def3}
\end{equation}
where $f(T_i)$ represents the set of text items associated with the technique. 

\subsection{System Overview}

\begin{figure*}[!t]
\centering
    \includegraphics[width=\linewidth]{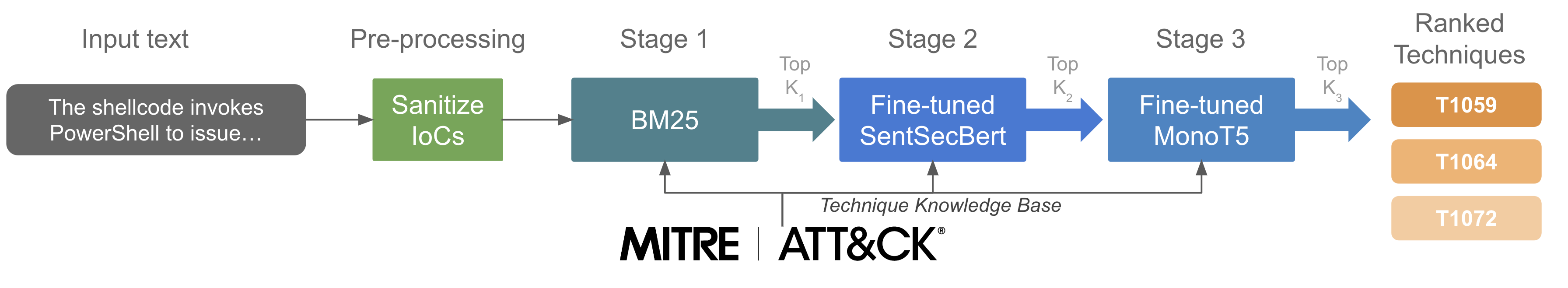}
    \caption{System Architecture of the Proposed Method ($K_i$ represents the number of candidate technique IDs at the output of the $i$-th stage, with  $K_1 > K_2 > K_3$).}
    \Description{Diagram of the high-level system components.}
\label{fig:components}
\end{figure*}

Our technique employs a multi-stage ranking architecture, which integrates multiple ranking models to achieve an optimal balance between effectiveness and efficiency.
In the multi-stage ranking approach, candidates identified by each stage become inputs for subsequent ranking stages. As depicted in Figure (\ref{fig:components}) the supervision level increases as we progress in the ranking pipeline. Hence we prevent over-fitting and ensure efficiency in the early stages. Stage 1 is fully unsupervised and allows for a fast filtering of candidate techniques using BM25. Stage 2 injects supervision via representation learning of the embedding space. This is achieved by fine-tuned SentSecBert. Stage 3 formulates the ranking as a supervised problem where relevance of a candidate text to a query is assessed via a classifier (Fined-tuned MonoT5). The softmax output of the classifier is used a ranking metric. Each stage receives a ranked list and produces a shorter ranked list by retaining only highly ranked candidates. 
That is, Eq. (\ref{eq:Def3}) is independently computed by each ranker on an iteratively reduced corpus $\mathcal{C}$.
The final output of the multi-stage ranking architecture is generated by the ranked list from the last stage.

Figure (\ref{fig:components}) provides an illustration of the components in the ranking pipeline of our system.
Before performing the ranking, both the text in the corpus and the query undergo preprocessing to remove less informative text elements, such as indicators of compromise. These indicators include \emph{file paths}, \emph{domains}, \emph{URLs}, \emph{email addresses}, \emph{IP addresses}, \emph{ASNs}, \emph{CVEs}, \emph{MITRE ATT\&CK techniques}, \emph{hashes}, \emph{registry keys} and \emph{Bitcoin addresses}. In the first stage, ranking is performed based on a bag-of-words representation, evaluating the exact match of keywords in the query with the text items in the corpus. To accomplish this, we create an inverted index and generate term frequency weighted document vectors.
Conversely, the following stages utilize and compare dense representations obtained through fine-tuning pre-trained LMs. The second stage employs a bi-encoder architecture following the formulation of Eq. (\ref{eq:Def1}), while the subsequent stage employs the computationally more intensive mono-encoder, Eq. (\ref{eq:Def2}), that allows interaction between the query and the candidate text items to finalize the rankings.

\section{Ranking Pipeline}

\subsection{Query Preprocessing}

Threat behaviors offer detailed descriptions of how threat actors implement various techniques during cyberattacks.
Specifically, they identify interactions among system entities (like files, processes, network sockets), communication patterns, network endpoints, code excerpts, commands, and the names of tools used in the attack. 
The terminology employed in these descriptions is domain-specific, which substantially differs from the natural text commonly used to train language models.
This discrepancy poses challenges when processing threat intelligence reports using language models.
A notable difficulty arises from the tokenization process applied by these models, wherein the text input is broken down into smaller pieces based on a predefined vocabulary that considers the occurrence frequency of text and the presence of common delimiters.
These tokenizers also split the commonly appearing and widely varying indicators of compromise (e.g., IP addresses, MAC addresses, file hashes, URLs, filenames, domain names, ASNs) into less informative pieces, making it harder for the model to properly contextualize such information. This becomes a primary concern when associating the query text extracted from threat intelligence reports with technique descriptions in the ATT\&CK knowledge base, as the technique descriptions are quite general.
To address this issue, we exploit the fact that many IoCs have a well-defined form and employ regular expression-based rules to identify and substitute these indicators of compromise with generic placeholder terms, such as "\textit{ip address}", "\textit{hash}", "\textit{URL}" etc., in our approach. 
For this, we leverage \textsc{iocparser}\footnote{\url{https://iocparser.com}} tool to identify all IoCs.


\subsection{Stage-1: Exact Term Matching}

The initial ranking stage of our ranking pipeline aims to enhance the efficiency and accuracy of subsequent ranking stages by eliminating some of the unrelated techniques. This is particularly crucial when considering the large number of techniques and (sub-)techniques, i.e., $|\mathcal{T}|$. 
To achieve this, we adopt a traditional information retrieval (IR) approach to retrieve the most relevant text item in our corpus based on exact term matching.
That is, terms from each $\textcolor{black}{it}\in\mathcal{C}$ and terms from queries had to match exactly to contribute to a relevance score.
This matching is typically performed using a normalized version of the terms obtained by stripping off inflectional parts (prefixes and suffixes) of the words.

To achieve this goal, we utilize the widely-adopted BM25 \cite{robertson1995okapi, crestani1998document} method, which has been a prevalent token-based matching algorithm to date \cite{tang2019distilling, lee2019latent, yates2021pretrained}.
The BM25 algorithm employs a high-dimensional sparse-vector representation that calculates the weight of each term based on its term frequency and inverse document frequency.
The BM25 relevance score between each $ti\in\mathcal{C}$ and the query $q$ is determined through inner products of their corresponding vectors, and the retrieval can be performed efficiently using an inverted index.
In our approach, we utilize the resulting score values to retain the top 100 most relevant techniques,
\textcolor{black}{as the primary objective of the first stage is to filter out potentially unrelated techniques.} 

\subsection{Stage-2: Transformer-Based Bi-Encoders for Semantic Matching}
In this stage, we exploit the language understanding capability of transformer-based deep neural networks.
It has been demonstrated that both small and large language models\footnote{
While there is no formal definition for what constitutes a small and large language model, we classify pretrained LMs with parameters up to several hundreds of millions, such as BERT \cite{devlin2018bert}, RoBERTa \cite{liu2019roberta}, and ERNIE \cite{sun2019ernie}, as small LMs. Conversely, models with more than billions of parameters, like closed GPT family models \cite{brown2020language} and open models such as various fine-tuned variants of LLAMA \cite{touvron2023llama}, are categorized as large LMs.} can very effectively capture complex
syntactical and semantic structures as well as how meanings vary across different linguistic contexts.
To retrieve the techniques that are semantically most relevant to a query, we use a small language model due to ease in their fine-tuning and deployment. 
(In Sec. \ref{subsec:comparison_llms}, we also compare our ranking system against well-known large LMs under a zero-shot test setting.)

Our ranker follows a bi-encoder architecture, where we independently compute dense-vector representations of text items, $\eta(it_i)$, and the query,  $\eta(q)$, as formulated in Eq. (\ref{eq:Def1}).
The relevance matching scores, $\phi(\eta(it_i), \eta(q))$, are computed in the resulting representation space by evaluating the cosine similarity between the embedded vectors.
For our encoding function, we adopt the architecture and hyper-parameters of the 12-layer BERT-base model. The BERT model is pretrained on a large corpus of English sentences using a masked language modeling task (MLM), where the objective is to predict the masked tokens in the input \cite{devlin2018bert}.
BERT has a maximum input length constraint of 512 (subword) tokens, approximately equivalent to 400 words, represented as 768-dimensional vectors. 
The model's output includes the same number of tokens, along with a special token, denoted as \texttt{[CLS]}, that aggregates the information in the entire input sequence. 

Developing a ranker with BERT requires performing two steps. 
Firstly, the amount of security-related text in the data used to pretrain BERT is likely limited. 
Further, due to the distinctive characteristics of security text compared to other natural language \cite{satvat2021extractor}, the limited exposure to security domain knowledge hinders 
the model's ability to properly contextualize security concepts.
Secondly, the ranker must undergo appropriate fine-tuning to accurately calculate semantic relevance scores needed to perform the technique ID annotation task.

Adapting BERT to the cybersecurity domain requires additional pretraining on threat intelligence text.
Several attempts have been made to achieve this, including CyBERT \cite{ranade2021cybert}, SecBERT\footnote{\hyperlink{https://github.com/jackaduma/SecBERT}{https://github.com/jackaduma/SecBERT}}, and Attack-BERT\footnote{\hyperlink{https://huggingface.co/basel/ATTACK-BERT}{https://huggingface.co/basel/ATTACK-BERT}}.
Whereas the fine-tuning process involves exposing the further-pretrained model to task-specific examples to align the model's output with the correct notion of semantic relatedness. 
During fine-tuning, the two encoders, Eq. (\ref{eq:Def1}), are combined by computing the inner product of their outputs. 
This bi-encoder backbone is then trained in an end-to-end manner using a task-specific dataset to maximize the cosine similarity computed between the outputs of the two encoders, specifically the \texttt{CLS} tokens.

In our evaluation, we found that SecBERT outperformed other models for our specific task, so we chose to fine-tune it. 
During the fine-tuning process, the model is presented with an equal number of positive and negative pairs of text items and queries.
For positive pairs, the cosine similarity is set to one to indicate that the two texts are related, while for negative pairs, it is set to negative one to indicate a lack of relevance. This setup essentially turns the model into a binary classifier for decision-making.
To generate the necessary data for fine-tuning, we utilized the samples from our custom dataset (Sec. \ref{sec:dataset}), which includes behaviors extracted from threat intelligence reports along with corresponding technique ID annotations. Since our data only includes queries, we used the title of the technique ID along with the general description (to fill 512 tokens) to create positive pairs, while for negative pairs, we included descriptions of non-matching technique IDs. 
For this we used 70\% all threat behavior descriptions in our dataset.  
To enhance the fine-tuning data, we also leveraged a subset of datasets used to build the widely-used sentence-transformer model that focuses on identifying sentences with similar meanings \cite{reimers2019sentence}. 
(These datasets are listed in Appendix Sec. \ref{apx:secbert_datasets}, comprising close 2M pairs of text samples.)
By the end of fine-tuning, both encoders are updated, and they can be used to generate vector representations. We refer to the fine-tuned version of the further-pretrained BERT model as the \texttt{SentSecBERT} model.
For a more detailed discussion and comparison of the rankers considered for the second stage, please refer to Sec. \ref{sec:ablation}.

During inference, we utilize the cosine similarity between two text embeddings as our score to rank the most relevant technique IDs to a given query. 
Our corpus $\mathcal{C}$ at this stage includes text excerpts extracted from the ATT\&CK knowledge base.
The text items associated with each technique ID $T_i$, i.e., $f(T_i)$, are obtained by segmenting the available text under the the main headings, including main description, detections, mitigations, and procedure examples, into 512 tokens in an overlapping manner with a stride of 128 tokens.
The vector representation for each text item is precomputed and stored for fast matching.

\subsection{Stage-3: Transformer-Based Mono-Encoder for Semantic Matching}

At this stage, we employ a ranker that better leverages the attention mechanism of a transformer. To achieve this, we utilize a single encoder that processes both the query and a text item simultaneously. This design enables interactions between query tokens and text item tokens, allowing the model to capture deeper semantic relationships within the input sequence. In contrast, a bi-encoder structure contextualizes each piece of text independently, potentially limiting the model's ability to fully exploit the connections between the query and the text item.
Thus, the model's input sequence combines the query $q$ and one of $it_i\in\mathcal{C}$ and its output designates whether the two sub-sequences are related.

Essentially, this mono-encoder design treats text ranking as a binary classification problem over all text items in the corpus, using a metric indicating the model's confidence as the ranking score. The advantage of this interaction-based ranking model is balanced by the higher latency it incurs, as the ranking computation must be performed at query time. However, since most unlikely techniques have already been eliminated in the first and second stages, it can be feasibly deployed on a smaller candidate pool. 
Further, since this model does not need to explicitly learn a dense vector representation, like performed by encoder-only transformer models, alternative architectures based on encoder-decoder architectures used can also be utilized for our purpose.
These generative models that are mainly designed for sequence-to-sequence to transformation shown to be quite effective in performing several tasks that do not necessarily involve a sequence generation \cite{raffel2020exploring, yang2020designing,pradeep2020scientific}. 

To fully leverage the language understanding capabilities, we developed our ranker using the T5 model, a pretrained sequence-to-sequence transformer \cite{raffel2020exploring}. 
Similar to the MLM task used for training BERT, T5's training objective involves generating a target text from a corrupted input text by predicting the missing parts after removing certain segments. With a larger number of parameters and training on a more extensive corpus, T5 provides enhanced language modeling capabilities compared to BERT \cite{nogueira2020document}.
Incorporating the T5 model as a ranker necessitates fine-tuning to ensure that, given a specific input template, the model generates a response suitable for use as a ranking score. This fine-tuning process is critical in guiding the model to produce the desired output when presented with inputs adhering to the predefined template.

To construct our ranker, we utilized the monoT5 model fine-tuned specifically for document ranking \cite{nogueira2020document}.
This fine-tuning was achieved using the MS-MARCO passage data, consisting of 8.8 million extracts from web pages in response to natural language questions from Bing search engine users \cite{nguyen2016ms}.
We further fine-tuned monoT5 to adapt it to our technique annotation task. 
For compatibility with the earlier fine-tuning process, we adopted the same input template defined as follows:\\
\centerline{\texttt{"Query:}[$q$] \texttt{Document:}[$it$] \texttt{Relevant:\{true,false\}}".}
Here, $q$ represents the query text, $it$ is a text item in $\mathcal{C}$, and the relevance between the two is indicated by either \texttt{true} or \texttt{false} words. 
Both of these words, \texttt{true} and \texttt{false}, are encoded as single tokens in T5's tokenization vocabulary. During model inference, the softmax function is exclusively applied to the logits of these two tokens, allowing us to determine their probabilities and make relevance assessments.
The ranking score of a text item in the corpus for a given query is then determined as the probability of the \texttt{true} token.

Similar to BERT, the T5 model is also limited to an input length of 512 tokens. Consequently, the combined length of $q$ and $it$ must not exceed this limit. To work within this constraint, we reserved 250 tokens for each query and text item. This length proves to be more than sufficient for text excerpts extracted from threat intelligence reports, as longer queries might include multiple threat behavior descriptions. 
To ensure each $it\in\mathcal{C}$ adheres to this length restriction, we restructured our corpus $\mathcal{C}$ by segmenting the information associated with each technique in the knowledge base into sequences of 250 tokens with a stride of 125 tokens. 
In the process of fine-tuning monoT5, we employed a subset of the query samples from our dataset (the same set utilized for constructing \texttt{SentSecBERT}) to generate an equal number of positive and negative $(q, it)$ pairs.
Each positive pair consists of $q$ and the first generated text item corresponding to ground truth technique ID, while a negative pair involves the same for a mismatching technique ID.

\section{Dataset Creation}
\label{sec:dataset}

The rankers used in the final two stages of our system are essentially language models fine-tuned to discern semantic relationships that are aligned with our task's objectives. This necessitates a substantial dataset of threat behaviors, each accompanied by its corresponding ground-truth adversarial technique IDs. We curate these threat behaviors from publicly available APT attack reports, many of which include varying degrees of technique annotations.
For example, in some cases, analysts will infer the techniques used in an APT attack, but without explicit references to specific sections of the text\footnote{Following are two examples of such reports:\\
https://cybersecurity.att.com/blogs/labs-research/crypto-miners-latest-techniques\\
https://www.trendmicro.com/en\_th/research/23/g/detecting-bpfdoor-backdoor-variants-abusing-bpf-filters.html}.
Hence, our focus was on threat reports in which analysts explicitly annotated relevant text or provided explanations for their annotation choices. We identified four such sources of reports information to compile the dataset for our task\footnote{Our combined dataset is publicly accessible at \href{https://github.com/qcri/Text2TTP}{https://github.com/qcri/Text2TTP}
}.

\begin{table}[!h]
\caption{Overview of Our Dataset}
\label{table:dataset_stats}
\begin{tabular}{lccc}
    \toprule
    \bf Source & \bf Techniques & \bf Behaviors & \bf Multi-Label \\
    \midrule
    ATT\&CK Reports & 92 & 2,017 & no \\
    CISA & 313 & 1,718 & yes \\
    TRAM & 137 & 1,506 & no \\
    WeLiveSecurity & 259 & 1,474 & no \\
   \midrule
   Combined Dataset & 410 & 6,612 & mixed \\
   \bottomrule
\end{tabular}
\end{table}

\paragraph{ATT\&CK Reports:}
Descriptions of techniques in ATT\&CK knowledge base often encompass lists of procedural examples that elucidate how specific APTs have employed that technique. Each procedure is accompanied by a brief summary and a reference to the source report. We used these summaries to manually pinpoint the corresponding text sections within the original report. Our team cross-referenced 1,180 summary procedure examples spanning 662 reports with the respective text passages in the original reports. Each identification was verified by another team member. When designation ambiguity arose, those techniques were omitted from the dataset.

\paragraph{WeLiveSecurity:}
WeLiveSecurity is a threat intelligence vendor that shares the APT reports created by its analysts on its website\footnote{\hyperlink{https://www.welivesecurity.com/en/}{https://www.welivesecurity.com/en/}}. Each published APT report includes a list of associated adversary techniques along with a detailed summary of observed threat behaviors at its conclusion. While the reports lack annotations at the passage level, the provided summaries contain an adequate level of detail for our task. As a result, we extracted a total of 1,474 threat behaviors from 101 reports.

\paragraph{CISA:}
The threat reports published by the Cybersecurity and Infrastructure Security Agency (CISA)\footnote{\hyperlink{https://www.cisa.gov/}{https://www.cisa.gov/}} provide summarized insights into observed APT attacks, offering a concise overview instead of exhaustive details. 
Each report contains sentence-level annotations of technique IDs, and in certain cases, some sentences carry multiple annotations at a sub-sentence level.
We noticed that in many instances, these succinct descriptions lack the depth required for comprehensive technique representations. Consequently, instead of segmenting these sentences into individual parts, we opted to annotate the complete sentence with multiple technique IDs. 
This means that these sentences are considered multi-labeled and are retrieved multiple times, each for the corresponding adversarial technique.
Our dataset comprises a total of 1,718 annotations extracted from 65 reports.

\paragraph{TRAM Training Data:}
TRAM, an open-source tool developed by the MITRE ATT\&CK team, is designed to enhance the efficiency of report annotation\footnote{\hyperlink{https://github.com/center-for-threat-informed-defense/tram/}{https://github.com/center-for-threat-informed-defense/tram/}}. It employs multiple machine learning models, including logistic regression, Naive Bayes, and multilayer perceptron, to classify behaviors and assign them to corresponding technique IDs. 
\textcolor{black}{To facilitate the training of these models, the TRAM source also includes a training dataset that includes an array of threat behavior descriptions of varying lengths, combined with technique ID data.
Our dataset has been augmented with this training data, encompassing 1,506 behavior descriptions across 137 techniques.
It is important to note that while TRAM is an annotation tool developed by the MITRE team, our dataset creation did not incorporate TRAM's annotations as supplementary data.}

Table \ref{table:dataset_stats} presents an overview of our dataset, specifying the count of covered adversarial techniques along with the quantity of threat behavior descriptions and their corresponding sources. These descriptions of threat behaviors vary in length, spanning from compact passages to succinct statements. 
\textcolor{black}{The combined dataset has 6,612 unique threat behavior descriptions aggregated from individual datasets. 
Within this collection, we identified an overlap of 103 threat behaviors across different datasets.}
\textcolor{black}{
It must be noted that the coverage of the number of distinct techniques varies across datasets. Specifically, the CISA dataset features 232 out of 313 techniques that are also found in other datasets. The WeLiveSecurity dataset has 220 techniques in common with other datasets. There is an overlap of 123 techniques in the TRAM dataset with others. Meanwhile, he ATT\&CK Reports dataset shares 73 techniques with other datasets.}


\begin{table*}[tbh]
\caption{\centering Performance of Proposed Multi-Stage Ranking Solution}
\label{table:res_pipeline}
\begin{tabular}{lcccccccc}
    \toprule
    \bf Stage & \bf Recall@100 & \bf Recall@50 & \bf Recall@10 & \textcolor{black}{\bf Recall@5} &\bf Recall@3 & \bf Precision@1 & \bf MRR\\
    \midrule
    Stage-1 (BM25) & 0.9736 & 0.9414 & 0.8069 & \textcolor{black}{0.7211} &0.6452 & 0.4702 & 0.5848 \\
    \textcolor{black}{S}tage-2 (SentSecBert) & - & 0.9595 & 0.8382 & \textcolor{black}{0.7359} & 0.6732 & 0.4529 & 0.5890 \\
    \textcolor{black}{S}tage-3 (MonoT5) & - & - & 0.9207 & \textcolor{black}{0.8671} & 0.8102 & 0.5783 & 0.7068 \\
    \bottomrule
\end{tabular}
\end{table*}

\section{Evaluation}
\label{sec:evaluation}

\subsection{Evaluation Setup}

Our experiments were conducted within the Ubuntu 18.04.2 operating system environment, utilizing an Intel Xeon Gold 6140 CPU with 72 cores, a Tesla V100-SXM2 32GB graphics card, and 504 GB of memory. 

\subsection{End-to-End Evaluation}

We assess the effectiveness of the proposed multi-stage ranking architecture. 
In our tests, the corpus $\mathcal{C}$ included descriptive text associated with 193 top-level techniques. While the threat behavior descriptions in our dataset only spanned 168 techniques, our aim was to avoid overestimating our system performance.
Out of the 6,612 samples in our dataset, 4,749 were used for fine-tuning Stage-2 and Stage-3 rankers and 1,212 samples related to 168 techniques were used during testing.
It's important to highlight that the ATT\&CK Reports dataset is generated through manual identification of the original descriptions linked to sample procedure examples found in the knowledge base within the relevant source reports. 
When testing samples from this dataset, we excluded all text items containing the corresponding procedure examples from our corpus to avoid a bias in our performance results. (That is, no threat behavior extracted from a report was compared against its summarized version in the knowledge base.)

\begin{figure}[!ht]
\centering
    \includegraphics[width=0.75\linewidth]{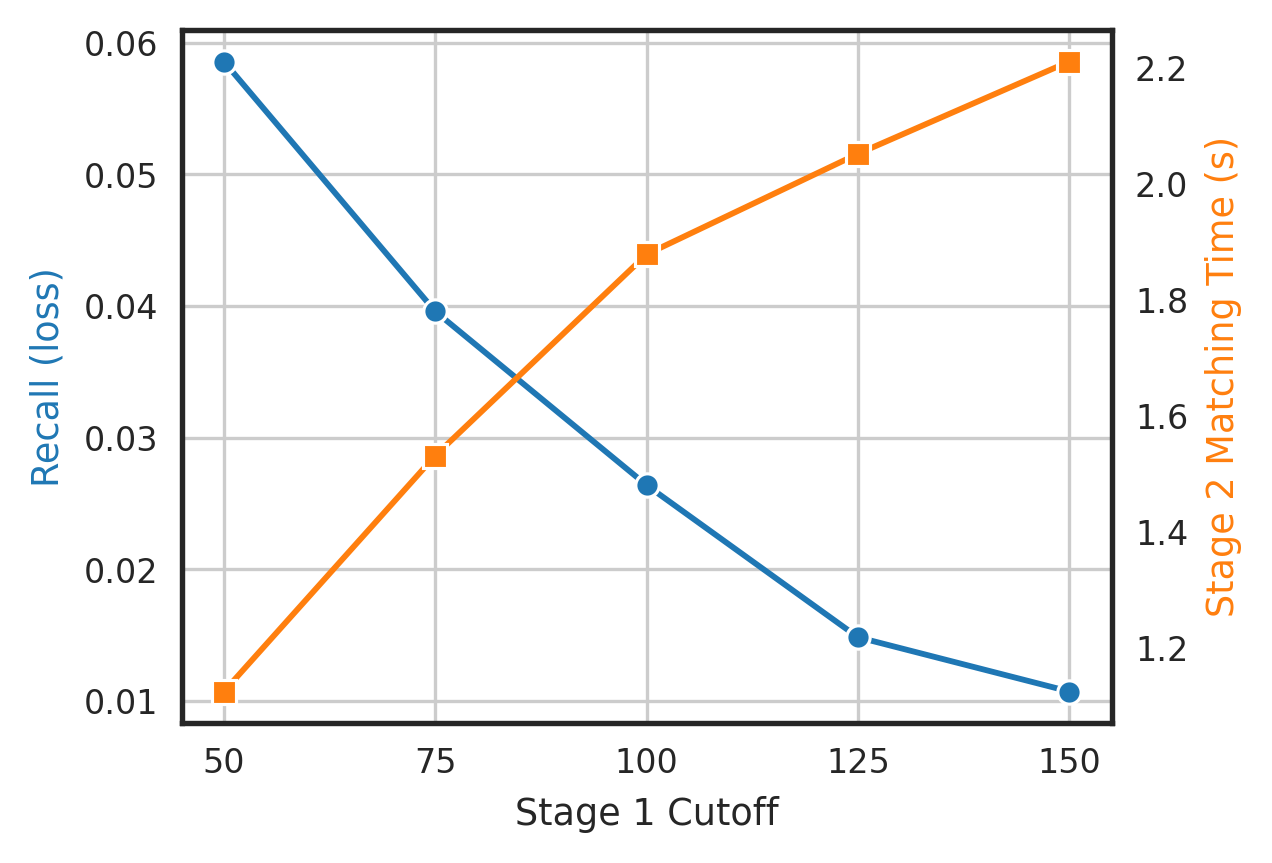}
    \caption{\textcolor{black}{The effect of raising the cutoff threshold from 50 to 150 in the Stage-1 ranker on the Stage-2 ranker's performance is depicted. The blue curve represents the decrease in recall performance, while the yellow curve illustrates the
    increase in computational time at the Stage-2 ranker for processing each query behavior.}}
\label{fig:stage_1_cutoff}
\end{figure}

The results obtained at different stages of our pipeline are presented in Table \ref{table:res_pipeline}, showcasing ranking performance across various metrics such as recall at top $K$ candidates, precision for the best guess, and the mean reciprocal rank (MRR).
The retrieval performance of each ranker used in distinct stages is shown in separate rows. For instance, when the initial stage ranker is confined to 100 candidates and the Stage-2 ranker to 50 candidates, the Stage-3 ranker achieves a remarkable recall@10 performance of 92\% and recall@3 performance of 81\% in accurately identifying the correct technique ID within its top candidates among more than 193 techniques in the ATT\&CK knowledge base.
It's important to note that the recall@100 values for Stage-2 and both recall@100 and recall@50 values for Stage-3 are not displayed, since the number of candidate techniques has been diminished by the preceding rankers.
\textcolor{black}{
Figure \ref{fig:stage_1_cutoff} demonstrates the trade-off between annotation accuracy and time complexity in our multi-stage ranker, showing the impact of adjusting the cutoff threshold in the Stage-1 ranker on recall performance and matching latency in the Stage-2 ranker.
}

These findings also illustrate that each successive stage enhances the system's proficiency in associating accurate MITRE ATT\&CK techniques with the queries.
To gauge the individual contribution of each ranker to the overall performance, we can assess the values in the Recall@10 and Recall@3 columns. We observe that BM25 scoring, utilizing a bag-of-words representation, achieves an independent recall performance of 80\% and 64\%, respectively. When combined with a bi-encoder-based semantic retriever in the two-stage ranker, performance further improves by over 3\%. Incorporating the mono-encoder-based semantic retriever further boosts the ranking performance by 9\% for Recall@10 and 4\% for Recall@3.

\subsection{Comparison to Previous Studies}

To better assess our approach's effectiveness, we compare our system's performance against three other state-of-the-art methods designed for annotating threat reports with adversarial technique IDs.
Among these, TRAM~\cite{tram:2023} employs a multi-class classification approach with classification confidence serving as the ranking score, while Ladder~\cite{ladder:alam:2022} and AttacKG \cite{li2022attackg} utilize sentence embeddings to measure similarity.
The latter two techniques are the most recent approaches to technique annotation task, and they are most similar to our technique in their approach to text representation. 
(Further details of these techniques are discussed in Sec. \ref{sec:otherwork}).
For our evaluation, we leveraged the publicly available source codes of these techniques.
Since TRAM supports several classification models, we built it from its source while incorporating all classification models.
Our findings revealed that the logistic regression classifier outperforms other models, hence we adopted it for our evaluation.

The tests were conducted on the same 1,212 threat behaviors used for the results of tests presented in Table \ref{table:res_pipeline}.
The comparison results are presented in Table \ref{table:compare_related_rank}.
These findings illustrate that our multi-stage ranking approach significantly outperforms the other methods by a considerable margin (+30\%) across all ranking configurations. For the best prediction scenario (Precision@1), our technique achieves an accuracy of 57\%, while the second-best performing method, TRAM, reaches 32\%.
In terms of F1-Macro and F1-weighted metrics, LADDER outperforms both TRAM and AttacKG but lags behind our method by a significant margin of approximately 15

\begin{table*}[!h]
\caption{\centering Comparison to  Related Work}
\label{table:compare_related_rank}
\begin{tabular}{ccccc|ccc}
    \toprule
    \bf Method & \bf \small{Recall@10} & \bf \small{Recall@5} & \bf \small{Recall@3}  & \bf \small{MRR}& \bf \small{Precision}@1 & \bf F1-Macro & \bf F1-Weighted\\
    \midrule
    TRAM~\cite{tram:2023} & 0.4582 & 0.4308 & 0.4028  & 0.3769 & 0.3204 & 0.1179 & 0.2357\\ 
    Ladder~\cite{ladder:alam:2022} & 0.6342 & 0.5198 & 0.4435  & 0.3985 & 0.2848 & 0.2256 & 0.2779\\
    AttacKG~\cite{li2022attackg} & 0.2454 & 0.1731 & 0.1261 & 0.1254 & 0.0593 & 0.0059 & 0.0258 \\
    Ours  & $\mathbf{0.9207}$ & $\mathbf{0.8671}$ & $\mathbf{0.8102}$ & $\mathbf{0.7068}$ & $\mathbf{0.5784}$ & $\mathbf{0.3664}$ & $\mathbf{0.4239}$\\
    \bottomrule
\end{tabular}
\end{table*}

\subsection{Comparison to LLMs}
\label{subsec:comparison_llms}


In this section, we explore how LLMs can accomplish treat report annotation task. Subsequently, we compare the effectiveness and execution time of our solution with those of large language models.

\subsubsection{Task description.} 
\label{subsubsec:task_description}
In this experiment, we prompt LLMs to identify the top-3 most likely MITRE ATT\&CK Technique IDs that an input threat behavior describes.

\subsubsection{Configuration}
\label{subsubsec:llm_configuration}
LLMs exhibit high sensitivity to both the formulation and structure of prompts, as well as the parameters employed. Hence, it holds significant importance to craft precise prompts that align with the specific task. In our scenario, the identification of technique IDs from descriptions hinges heavily on the LLM's underlying training data. 


\paragraph{Prompt Engineering:} LLMs are trained with specific predefined prompt structures, unique to each model. While common templates can be shared among some LLMs, optimal results are achieved by using the exact prompt template used during the LLM's training. These templates consist of four components: Instruction (task for the LLM), Context (external information aiding the LLM), Input Data (the query), and Output Indicator (desired output format). In our case, we employ the relevant template for each LLM used.

In addition, we utilize various effective prompting techniques to enhance LLM performance: \textit{Role Play}, \textit{Specificity and Precision},
and \textit{ReAct Prompting}, and \textit{Few-shot Prompting}.
\textit{Role Play} prompts elicit improved performance from LLMs when they adopt specific roles aligned with the task. 
In our case, each prompt is introduced with the phrase \texttt{Act as an experienced security analyst, [...]}, assigning a distinct role to guide the LLM's response.
\textit{Specificity and Precision} in prompts indicate that LLMs achieve improved results when the prompts are precise and well-formatted.
In our scenario, we instruct the LLM to provide answers in a well-defined JSON format. By specifying the format and appropriate parameters, we avoid hallucinations and ensure that the generated text remains relevant to the assigned task.
\textit{ReAct \cite{yao2022react} Prompting} is designed to compel LLMs to generate reasoning traces and task-specific actions.
We employ it to produce reasoning traces for the answers it generates by asking for an explanation of why the LLM chose specific technique IDs for a given sentence. 
This step also helps in reducing hallucinations and improving the quality of results. 
\textcolor{black}{We demonstrate the application of these prompting techniques in our prompt used in the zero-shot learning setting, as shown in Fig. \ref{fig:prompt} in Appendix Sec. \ref{apdx:prompts}.}


\textcolor{black}{\textit{Few-shot Prompting:} This is a prompting technique that helps LLMs in generating coherent and contextually appropriate responses with only a limited amount of input examples. 
These examples act as a context for the model to infer the underlying pattern or classification rule.
Given that our study involves over 100 technique categories, we limited our consideration to a maximum of 20 examples.
Our selection of examples for the few-shot approach was guided by the data distribution. Specifically, we chose 40\% of the examples from techniques with more than 30 instances, another 40\% from techniques having 6 to 30 instances, and the remaining 20\% from techniques with 1 to 5 instances.
For this purpose, we select two techniques that are semantically most similar to the correct one, treating them as potential correct choices. Then, we take the description of the correct technique and integrate it into the explanation.
We show a sample prompt with two examples in Fig. \ref{fig:few_shot_5}.
}





\paragraph{Parameters:} LLMs offer several parameters that can be tuned to optimize their performance for specific tasks, mainly \textit{temperature, top-k, top-p}, and \textit{maximum new tokens}.
The \textit{temperature} parameter controls the randomness of token generation during text generation. It affects how the model selects the next token based on the predicted probabilities of the possible tokens.
The \textit{top-k} parameter instructs the model to select the next token from the top 'k' tokens in its sorted list based on their probabilities. 
The \textit{top-p} parameter of a language model (LLM) controls the dynamic sampling of tokens during text generation.
When using \textit{top-p}, the model samples  tokens among the candidates with the cumulative probability surpassing the specified threshold $p$.
The \textit{maximum new tokens} parameter limits the length of the generated output. 
In our experiment, we set the temperature to 0.7, top-p to 0.60, top-k to 40, and maximum new tokens to 200. 

\subsubsection{Models}
\label{subsubsec:llm_models}

In this work, we used a combination of open models from HuggingFace\footnote{\url{https://huggingface.co}} and closed source models from OpenAI\footnote{\url{https://openai.com/gpt-4}}.
For the open models, we selected those that could run on a single NVIDIA A100-80GB GPU.
\textcolor{black}{Our model selection criterion was guided by their performance on the Chatbot Arena Leaderboard\footnote{https://huggingface.co/spaces/lmsys/chatbot-arena-leaderboard}, as further detailed in \cite{zheng2023judging}. 
We selected the top-performing model in each size category, specifically those with 7B, 13B, and 30B parameters.}
Table \ref{table:used_llms} displays the list of employed language models in our study. 

\begin{table}[!htb]
\caption{\centering Employed Large Language Models}
\label{table:used_llms}
\begin{tabular}{lc}
\toprule
\textbf{Model Name} & \textbf{HuggingFace Path}                \\ 
    \midrule
Vicuna-33B   \cite{zheng2023judging}      & lmsys/vicuna-33b-v1.3          \\ 
WizardLM-13B \cite{xu2023wizardlm}         & WizardLM/WizardLM-13B-V1.2      \\ 
Zephyr-7B \cite{tunstall2023zephyr}         & HuggingFaceH4/zephyr-7b-beta      \\ 
gpt-4-32k               & -                               \\
\bottomrule
\end{tabular}
\end{table}

\subsubsection{Results}
\label{subsubsec:llm_results}

\begin{table}[!htb]
\caption{\centering Comparison to LLMs}
\label{table:comparison_llms}
\begin{tabular}{lcc}
\toprule
\textbf{Model Name} & \textbf{recall@3} & \textbf{Avg. Time per Sentence (s)} \\ \midrule
Vicuna-33B         & 0.0158            & 13.28                  \\ 
WizardLM-13B           & 0.1198            & 15.73                    \\ 
Zephyr-7B           & 0.1081            & 11.28                    \\ 
gpt-4-32k               & 0.4406            & 9.66 \\
Ours                & \textbf{0.8102}   & \textbf{8.90}               \\ 
\bottomrule
\end{tabular}
\end{table}



We conducted queries using the same set of 1,212 test threat behavior descriptions across the LLMs. 

\textcolor{black}{\textbf{Zero-Shot Learning Performance.}} The corresponding recall@3 results for correctly identifying the technique ID among the three identified ones are presented in Table \ref{table:comparison_llms}. These findings indicate that open LLMs exhibit limited performance in this task, with recall@3 rates remaining below 4\% for all of them. However, GPT-4 shows relatively better performance, achieving a recall@3 score of 44.06\%. The low recall@3 achieved by open models is primarily a result of generating invalid technique IDs that are not found in the MITRE ATT\&CK knowledge base. 

\begin{figure}[!ht]
\centering
    \includegraphics[width=3.1in]{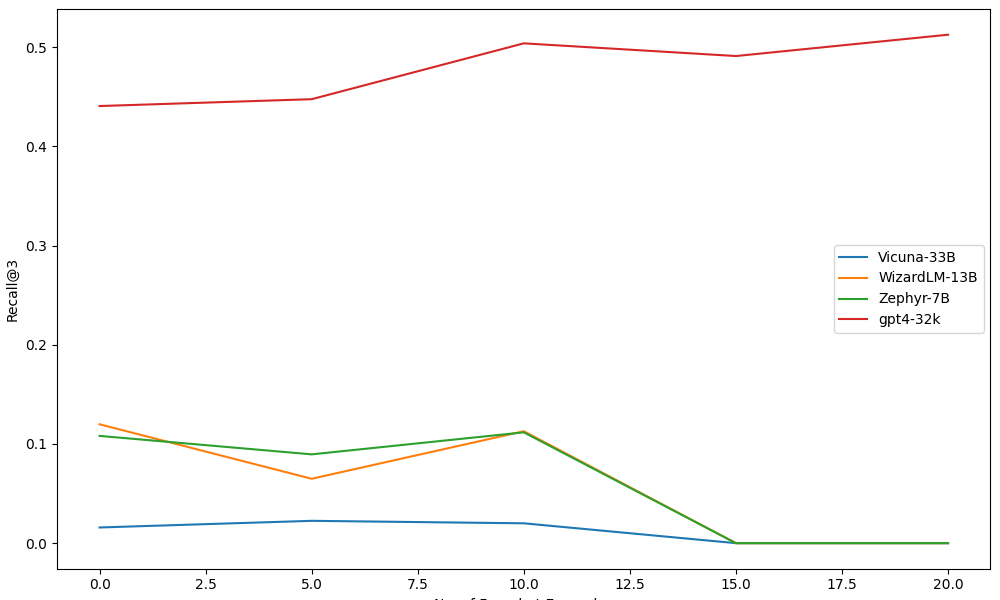}
    \caption{\textcolor{black}{Impact of increasing the number of few-shot learning examples on the measured Recall@3 score.}}
\label{fig:few_shot_examples}
\end{figure}

\textcolor{black}{
\textbf{Few-Shot Learning Performance.} 
Figure \ref{fig:few_shot_examples} illustrates the impact of adding different numbers of examples, each from a distinct technique, to our input prompt on the recall@3 score.
We observed that the few-shot learning setting does not enhance the performance of the open models, with the exception of a minor improvement in the Vicuna-33B model.
For the gpt4-32k model, the outcomes were nearly the same for 0 and 5 examples. Yet, there was a noticeable 5\% increase with the addition of 10 examples. Subsequently, the results remained steady, showing a slight decline of 1.2\% with 15 examples and a marginal rise of 0.87\% for 20 examples.
Remarkably, with the open LLMs, the recall@3 score decreases to 0 for all models after incorporating 15 examples.
Our investigation into this matter revealed that the LLMs tend to produce hallucinations, generating incomplete and arbitrary sentences drawn from the MITRE ATT\&CK knowledge base or techniques provided as part of the examples. Furthermore, they fail to comply with the given directives regarding specificity and precision
These results underscore the importance of fine-tuning these models on datasets related to technique annotation task to improve their retrieval accuracy, as we have done in our approach.
} 

\textbf{Inference time.} Despite the added network latency, our measurements reveal that the GPT-4 model,  presumed to possess the largest number of parameters, still manages a fairly prompt response time (9.66 seconds per query), which is comparable to our technique (8.9 seconds per query).

\textbf{Cost analysis.} \textcolor{black}{Running LLMs is a costly process. To rent an NVIDIA A100-80GB GPU, the cost in the market is around \$6.50\footnote{https://huggingface.co/pricing\#endpoints} per hour. On average, the experiments we run (i.e., with and without few-shot examples for all LLMs), lasted 8 hours. Thus, the cost of a single experiment is approximated to be \$48. In other words, on average the cost of a sentence query is approximately \$0.04.}
\textcolor{black}{
For GPT4, the experiments (with and without few-shot examples) cost \$749.99 in total. On average, the cost of a sentence query is approximately \$0.4. 
In the 0-shot learning experiment, the average cost per sentence is \$0.006, compared to \$0.1374 per sentence in the 20-shot experiment.
}

\subsection{Error Analysis}
\begin{figure}[ht]
\centering
\begin{subfigure}{\linewidth}
\includegraphics[width=\linewidth]{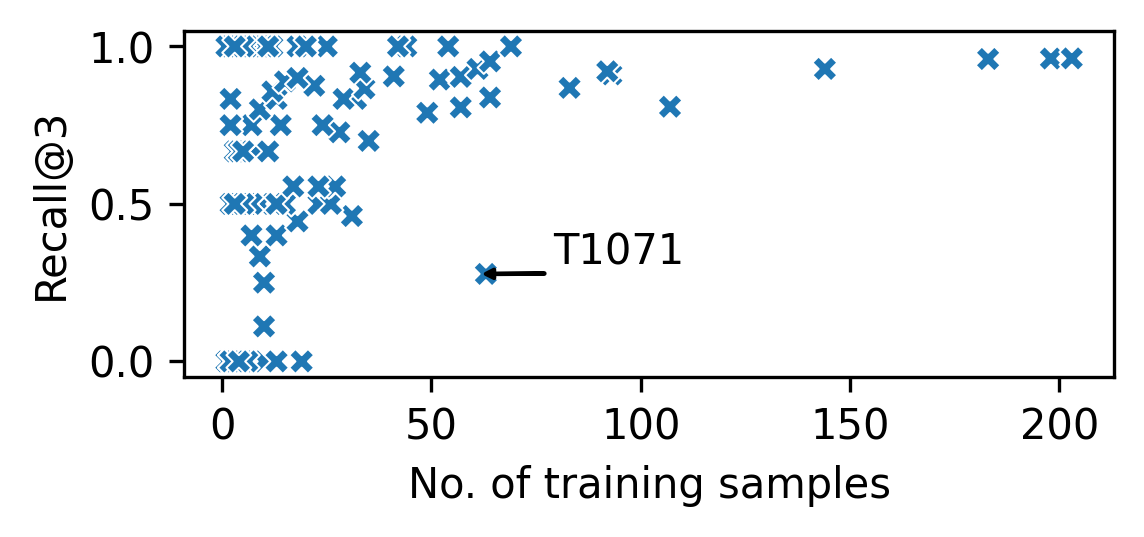} 
\caption{}
\label{fig:r_vs_train}
\end{subfigure}

\begin{subfigure}{\linewidth}
\includegraphics[width=\linewidth]{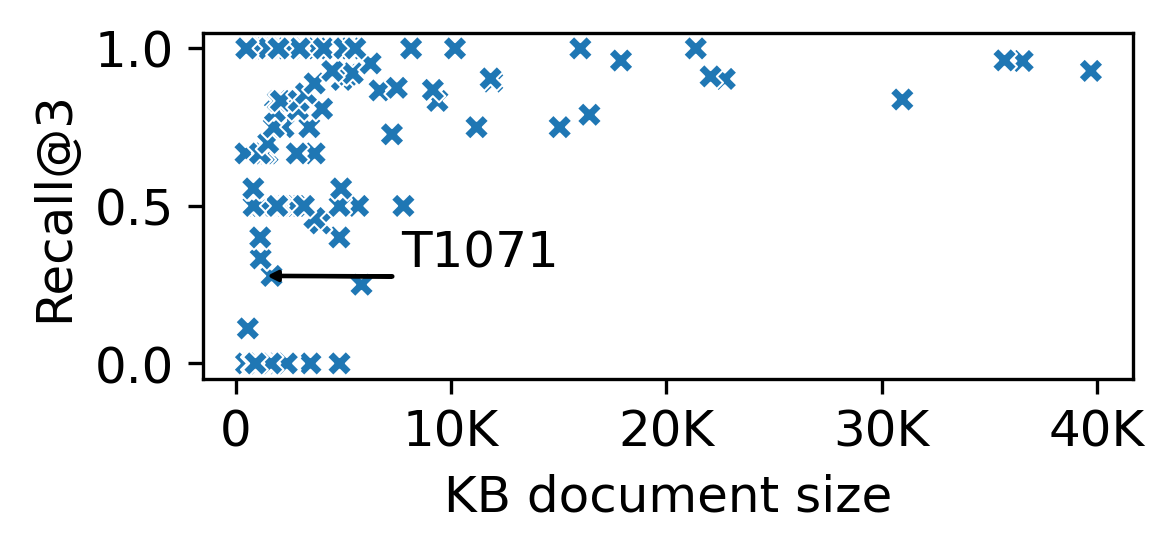}
\caption{}
\label{fig:r_vs_doc}
\end{subfigure}

\caption{\textcolor{black}{(a) Recall at top 3 vs the number of training samples available for each technique, (b) Recall at top 3 vs MITRE ATT\&CK document size (in characters) for each technique.}}
\label{fig:ranking_errors}
\end{figure}

\textcolor{black}{In our ranking pipeline, we've noticed that certain techniques are more susceptible to errors. This can be traced back to two main reasons. Firstly, the number of samples available for fine-tuning our Stage-2 and Stage-3 rankers for these specific techniques is limited. Consequently, our rankers have reduced exposure to these techniques, impairing their ability to accurately identify them when mentioned in reports. This limited exposure is often a result of the relatively rare application of these techniques. For example, the technique `T1218 - System Binary Proxy Execution' is referenced in just one report within the ATT\&CK knowledge base.
Secondly, there is a negative correlation between the volume of information in our corpus about a technique and the likelihood of erroneous predictions. That is, the more comprehensively a technique is described in the knowledge base, the better our system is at distinguishing it from others. For instance, the technique `T1071 - Application Layer Protocol' has a very brief description in the knowledge base, leading to incorrect identifications by our system. 
Figure 7 displays the relation between recall misses and the number of samples employed during fine-tuning, as well as the volume of information available for each technique in our corpus. Each point in the figure represents a distinct technique
In Fig. \ref{fig:r_vs_train}, technique 'T1071' appears as an outlier in the plot of the technique’s recall in the top 3 versus the number of training samples. However, Fig. \ref{fig:r_vs_doc} explains this anomaly, indicating that technique 'T1071' is described very briefly in the ATT\&CK knowledge base.}

\textcolor{black}{
Regarding LLMs, we notice that these models generate new technique titles and associate them with technique IDs in their explanations. In addition, when adding more than 15 few-shot examples to the prompt, we notice that the LLMs loose context and do not follow the instructions. They hallucinate and generate random sentences by either repeating some examples or formulating new sentence from MITRE ATT\&CK knowledge base.
This underscores the need for background knowledge that these variants of open LLMs have not been trained on. While this limitation could potentially be mitigated by fine-tuning these LLMs specifically for this task, it would require substantial computational resources and training data.
}

\subsection{Ablation Analysis}
\label{sec:ablation}
We delve into the design considerations behind the rankers deployed in the last two stages of our system. For Stage-2, we've embraced a bi-encoder architecture, to create representations of our corpus derived from the ATT\&CK knowledge base and the query behaviors from our dataset.
The most eminent model for evaluating text similarity is the sentence transformers \cite{sbert:corr:2019}, denoted as SBERT. This model is generated through fine-tuning BERT on a generic sentence similarity evaluation task. 
In contrast, our Stage-2 ranker relies on a further pretrained version of BERT on 5K APT attack reports to adapt it to the cybersecurity domain, followed by a fine-tuning process to capture semantic relevance.  

An alternative to SentSecBERT is the publicly available Attack-BERT model, which is designed based on a BERT variant primarily trained to map \textit{common vulnerabilities and exposures} (CVE) entries to ATT\&CK techniques using textual similarity \cite{abdeen2023smet}.
A comparison involving the substitution of the Stage-2 ranker with these four models is presented in Table \ref{table:ablation_bi_enc}.
Notably, the fine-tuned SentSecBERT consistently outperforms the other models by a margin of at least 1.5\%. 
We also assessed the impact of fine-tuning the mono-encoder architecture, specifically monoT5, on performance improvement.
We also assessed how fine-tuning the mono-encoder architecture, i.e., the monoT5, improves the performance.
The outcomes of replacing the Stage-3 ranker with monoT5 are detailed in Table \ref{table:ablation_cross_enc}, revealing a potential enhancement of up to 9\%.

\begin{table}[!htb]
\caption{\centering Ablation Analysis of Stage-2 Ranker}
\small
\label{table:ablation_bi_enc}
\begin{tabular}{lcccc}
    \toprule
    \bf Model & \bf Recall@50 & \bf Recall@20 & \bf Recall@10 & \bf Recall@3 \\
    \midrule
    SBert~\cite{sbert:corr:2019} & 0.9505 & 0.8969 & 0.8251 & 0.6576 \\
    Attack-Bert\footnote{\url{https://huggingface.co/basel/ATTACK-BERT}} & 0.9414 & 0.8903 & 0.8350 & 0.6535 \\
    SentSecBert & 0.9554 & 0.8903 & 0.8226 & 0.6394 \\
    FT SentSecBert  & $\mathbf{0.9596}$ & $\mathbf{0.9059}$ & $\mathbf{0.8383}$ & $\mathbf{0.6732}$ \\
    \bottomrule
\end{tabular}
\end{table}

\begin{table}[!htb]
\caption{\centering Ablation Analysis of Stage-3 Ranker}
\label{table:ablation_cross_enc}
\begin{tabular}{lccc}
    \toprule
    \bf Model & \bf Recall@20 & \bf Recall@10 & \bf Recall@3 \\
    \midrule
    MonoT5 & 0.9101 & 0.8564 & 0.7244 \\
    FT MonoT5  & $\mathbf{0.9480}$ & $\mathbf{0.9208}$ & $\mathbf{0.8102}$ \\
    \bottomrule
\end{tabular}
\end{table}

In our fine-tuning process for the rankers, we've utilized 70\% of the threat behavior descriptions in our dataset, combining information from four distinct sources as outlined in Table \ref{table:dataset_stats}. This prompts the consideration of how using a singular dataset for fine-tuning rankers impacts their performance when tested on the other three.
To address this, we fine-tuned the model on one dataset and evaluated it on the remaining three datasets.
The performance outcomes for the Stage-2 and Stage-3 rankers are displayed in Table \ref{table:cross_ds_sentsecbert} and \ref{table:cross_ds_monot5}, respectively.
Interestingly, these results demonstrate that the optimal test performance is achieved by fine-tuning the model using manually extracted behaviors from ATT\&CK Reports.
Conversely, the least favorable test performance arises from rankers trained on the CISA dataset, likely due to its succinct descriptions that encompass multiple behaviors.
It must be emphasized here that the combined dataset offers the broadest technique coverage. 
Moreover, these variations in analysts' writing styles also reflect the real-world diversity in describing threat behaviors.
These suggest that individual datasets shouldn't be employed for training the multi-stage ranker.

\begin{table}[!h]
\footnotesize
\caption{Impact of Using Different Datasets for Fine-Tuning Stage-2 Ranker}
\label{table:cross_ds_sentsecbert}
\begin{tabular}{llccc}
    \toprule
    \bf Test & \bf Train & \bf Recall@50 & \bf Recall@20 & \bf Recall@10 \\
    \midrule
    \multirow{3}{*}{CISA} & ATT\&CK Rpts. & 0.8967 & 0.8040 & 0.7124 \\ 
    & TRAM & 0.8932 & 0.7899 & 0.6995 \\ 
    &WeLiveSecurity & 0.9026 & 0.7934 & 0.7183 \\ 
    \midrule
    \multirow{3}{*}{WeLiveSecurity} & ATT\&CK Rpts. & 0.9649 & 0.9263 & 0.8889 \\ 
    & TRAM & $\mathbf{0.9705}$ & 0.9331 & 0.8844 \\ 
    & CISA & 0.9694 & $\mathbf{0.9456}$ & 0.$\mathbf{9138}$ \\     
    \midrule
    \multirow{3}{*}{ATT\&CK Rpts.} & TRAM & 0.9842 & 0.9339 & 0.8746 \\ 
    & WeLiveSecurity & 0.9887 & 0.9422 & 0.8844 \\ 
    & CISA & $\mathbf{0.9902}$ & $\mathbf{0.9557}$ & $\mathbf{0.9084}$ \\ 
    \midrule
    \multirow{3}{*}{TRAM} & ATT\&CK Rpts. & 0.9245 & 0.8455 & 0.7540 \\
    & WeLiveSecurity & 0.9307 & 0.8641 & 0.7638 \\ 
    & CISA & $\mathbf{0.9405}$ & $\mathbf{0.8783}$ & $\mathbf{0.7806}$ \\ 
    \bottomrule
\end{tabular}
\end{table}

\begin{table}[!h]
\caption{Impact of Using Different Datasets for Fine-Tuning Stage-3 Ranker}
\footnotesize
\label{table:cross_ds_monot5}
\begin{tabular}{llccc}
    \toprule
    \bf Test & \bf Train & \bf Recall@10 & \bf Recall@5 & \bf Recall@3 \\
    \midrule
    \multirow{3}{*}{CISA} & ATT\&CK Rpts. & 0.7300 & 0.6432 & 0.5481 \\ 
    & TRAM & 0.7042 & 0.6080 & 0.5059 \\ 
    & WeLiveSecurity & 0.6784 & 0.5469 & 0.4354 \\ 
    \midrule
    \multirow{3}{*}{WeLiveSecurity} & ATT\&CK Rpts. & 0.9297 & 0.8617 & 0.7891 \\ 
    & TRAM & 0.9127 & 0.8447 & 0.7642 \\ 
    & CISA & 0.9240 & 0.8639 & 0.8027 \\ 
    \midrule
    \multirow{3}{*}{ATT\&CK Rpts.} & TRAM & 0.9775 & 0.9580 & 0.9062 \\ 
    & WeLiveSecurity & 0.9610 & 0.8851 & 0.7590 \\ 
    & CISA & 0.9865 & 0.9700 & 0.9384 \\ 
    \midrule
    \multirow{3}{*}{TRAM} & ATT\&CK Rpts. & 0.7673 & 0.6599 & 0.5853 \\
    & WeLiveSecurity & 0.6856 & 0.5595 & 0.4272 \\ 
    & CISA & 0.7540 & 0.6545 & 0.5533 \\ 
    \bottomrule
\end{tabular}
\end{table}

\section{Discussion}
\label{sec:discussion}

\paragraph{Error Analysis:} 
We analyzed the sources of errors for both the Stage-2 and Stage-3 rankers, as presented in Table \ref{table:res_pipeline}. To achieve this, we classified the threat behaviors in our test split that are ranked within the top-$K$ category based on their source dataset. The performance of the Stage-2 ranker in correctly ranking behaviors across four datasets is demonstrated in Table \ref{table:eval_sentsecbert_ft}. The recall values at various top-$K$ values indicate that the model's performance varies depending on the source of the sentence. 
These results demonstrate that our multi-stage ranker performs relatively better on descriptions from ATT\&CK Reports and WeLiveSecurity datasets compared to the TRAM and CISA datasets.
The corresponding outcomes for the Stage-3 ranker are provided in Table \ref{table:eval_monot5_ft}. A similar pattern is observed, but with an amplified performance gap between the two groups.
This observation can be attributed to two primary factors. The first factor is the sentence structure itself. TRAM and CISA descriptions tend to be concise, while others are more verbose. The second factor is the imbalance in the coverage of ATT\&CK techniques across each dataset.

\begin{table}[!h]
\caption{\centering Stage-2 Ranker Performance Across Individual Datasets}
\label{table:eval_sentsecbert_ft}
\begin{tabular}{lccc}
    \toprule
    \bf Dataset & \bf Recall@50 & \bf Recall@20 & \bf Recall@10 \\
    \midrule
    CISA & 0.9156 & 0.8089 & 0.7200 \\ 
    TRAM & 0.9434 & 0.8994 & 0.8176 \\ 
    WeLiveSecurity & 0.9725 & 0.9529 & 0.9020 \\ 
    ATT\&CK Reports  & 0.9879 & 0.9348 & 0.8792 \\ 
    \bottomrule
\end{tabular}
\end{table}

\begin{table}[!h]
\caption{\centering Stage-3 Ranker Performance Across Individual Datasets}
\label{table:eval_monot5_ft}
\begin{tabular}{lccc}
    \toprule
    \bf Dataset & \bf Recall@10 & \bf Recall@5 & \bf Recall@3 \\
    \midrule
    CISA & 0.8311 & 0.7511 & 0.6933 \\ 
    TRAM & 0.8616 & 0.7516 & 0.6792 \\ 
    WeLiveSecurity & 0.9686 & 0.9451 & 0.8902 \\ 
    ATT\&CK Reports & 0.9855 & 0.9710 & 0.9251 \\ 
    \bottomrule
\end{tabular}
\end{table}

\paragraph{Report Sentence Structures:} 
The threat behaviors in our dataset originate from various sources of threat reports, demonstrating a wide range of textual structures employed by analysts. 
The results of Tables \ref{table:cross_ds_sentsecbert}-\ref{table:cross_ds_monot5} and Tables  
\ref{table:eval_sentsecbert_ft}-\ref{table:eval_monot5_ft}.
also suggest that utilizing the TRAM dataset for fine-tuning or testing generally leads to the lowest performance.
This dataset predominantly consists of brief descriptions or phrases associated with attack activities. We posit that such succinct descriptions lacking adequate context may not provide an accurate basis for language models to perform semantic evaluation. 



\section{Related Work}
\label{sec:otherwork}
Gathering technique-level knowledge from threat analysis reports with reference to the ATT\&CK knowledge base has gained prominence due to the growing adoption of this framework as a point of reference by numerous cybersecurity detection and mitigation tools.
Therefore, several approaches have been proposed to automate the retrieval of ATT\&CK techniques from threat reports.  

Earlier approaches utilized conventional text representations for this purpose.
These included sparse representations based on bag-of-words features, including TF-IDF \cite{legoy2020automated, tsai2020cti} and frequencies of bi-grams and tri-grams \cite{legoy2020automated, tram:2023}, as well as dense word embeddings like Word2Vec \cite{legoy2020automated}.
These text representations were subsequently integrated with various classification models, considering both multi-class and multi-label classification scenarios.
These models were trained on datasets manually curated from threat intelligence reports. Following a comprehensive evaluation, \cite{legoy2020automated} determined that the combination of TF-IDF features with a linear SVM classifier yielded the most favorable performance.

The initial text representation approach lacked any specific alignment with the task and did not incorporate any relevant elements from it.
Therefore, subsequent methods expanded upon this approach to achieve a more nuanced understanding of threat behaviors. To this objective, \cite{ttpdrill:acsac:2017} proposed TTPDrill, a method that extracts threat actions from threat reports and aligns them with attack patterns.
This alignment is achieved through a predefined ontology that captures the relationships between various components of threat actions.
The ontology utilized in TTPDrill was primarily built upon MITRE's Common Attack Pattern Enumeration and Classification (CAPEC) catalog and the ATT\&CK threat repository. These resources offer a comprehensive compilation of attack patterns, tactics, techniques, and procedures that adversaries employ to exploit system vulnerabilities.
TTPrill employed part-of-speech parsing to identify candidate threat actions, in the form of $(subject, verb, object)$ tuples. These actions were then weighted and filtered using a BM25-based method to identify the best-matching threat actions within the threat ontology.

Another approach in this direction is the AttacKG method \cite{attackg:esorics:2022}. 
To identify adversarial techniques, AttacKG creates technique templates.
These are essentially attack graphs built upon technique procedure examples crawled from threat reports, consisting of attack entities as nodes and their interconnections as edges.
Attack entities include both subjects of attack behaviors and system-level objects they target.
This information is extracted using a regex-based IoC detector and a 
pretrained named entity recognition module.
AttacKG employs a graph alignment algorithm to map threat behaviors extracted from reports onto these predefined technique templates. The success of this method hinges on accurately identifying entities and their interdependencies, which is a distinct challenge in extraction of threat knowledge \cite{extractor:eurosp:2021,gao2021enabling,ji2022knowledge,ren2022cskg4apt}.

With the rising popularity of Transformer-based language models, recent approaches exploited their language modeling capabilities to implement semantic similarity-based approaches for the mapping task.
Using a pretrained sentence transformer model \cite{sbert:corr:2019}, these work compute the embeddings of the descriptions and evaluated relevance based on similarity of these embeddings. 
In this regard, TIM \cite{you2022tim} used two representations obtained from three-sentence long chunks of text from reports: a text embedding and a vector of values indicating the  
number of different types of IoCs observed in those passages. The similarities of these two vectors are then evaluated to ground truth threat behaviors extracted from other reports. 
In contrast to this approach, LADDER \cite{ladder:alam:2022}  used only attack phrases extracted from the relevant parts of the sentences identified to contain attack pattern descriptions and evaluated the similarity with respect to title and description of the ATT\&CK technique descriptions.

The most similar approaches to ours are the last two, \cite{you2022tim} and \cite{ladder:alam:2022}. In comparison to these works, we present a novel problem formulation through a multi-stage ranking scheme, which enables us to assess semantic similarity at varying levels. Importantly, our ranking models are constructed using language models that are tailored to the cybersecurity domain and fine-tuned for our specific task, resulting in improved accuracy in technique identification. Moreover, in our evaluation, we extensively explore the performance of several advanced LMs, trained on web-scale text corpora, for this particular task.

Another crucial aspect to consider is the collection of threat behavior descriptions needed for training and testing these approaches. 
Proposed methods have so far used custom-built datasets that vary in their coverage of adversarial techniques, ranging from as few as six techniques \cite{you2022tim} to as many as 179 techniques \cite{attackg:esorics:2022}.
Notably, the largest of these custom datasets contained around 7.3K technique descriptions \cite{attackg:esorics:2022}.
In this context, as part of our work we introduce a public dataset with 6.6K annotations spanning over 410 techniques.

\section{Conclusion}

We introduce a new method for extracting structured threat behaviors from threat intelligence text. Our method is based on a multi-stage ranking architecture that allows jointly optimizing for efficiency and effectiveness. 
Therefore, we believe this problem formulation better aligns with the real-world nature of the task considering the large number of adversary techniques and the extensive body of threat intelligence created by security analysts. Our findings show that the proposed system yields state-of-the-art performance results for this task. Results show that our method has a top-3 recall performance of 81\% in identifying the relevant technique among 193 top-level techniques.
Our tests also demonstrate that our system performs significantly better (+40\%) than the widely used large language models when tested under a zero-shot setting.


\bibliographystyle{ACM-Reference-Format}
\bibliography{references.bib}


\appendix

\section{Datasets Used for Fine-Tuning SecBert}
\label{apx:secbert_datasets}

\begin{table}[hb]
\caption{\centering Datasets Used for Fine-Tuning SecBert Model}
\label{table:ft-datasets}
\begin{tabular}{l|c}
    \toprule
    \bf Dataset & \bf \# text pairs \\
    \midrule
    specter\_train\_triples & 684,100 \\
    stackexchange\_duplicate\_questions\_title\_title & 304,525 \\
    AllNLI & 277,230 \\
    sentence-compression & 180,000 \\
    wikihow & 128,542 \\
    SimpleWiki & 102,225 \\
    NQ-train\_pairs & 100,231 \\
    squad\_pairs & 87,599 \\
    TriviaQA\_pairs & 73,346 \\
    \hline
    \bf Total & $\mathbf{1,937,798}$ \\
    \bottomrule
\end{tabular}
\end{table}

The pretrained SecBert model is fine-tuned to facilitate semantic text similarity, aligning with our technique annotation task's requirements. Due to the finite size of our threat behavior description dataset, we incorporated some of the \textcolor{black}{datasets\footnote{https://huggingface.co/datasets/sentence-transformers/embedding-training-data}} utilized in training the sentence transformer model as detailed in Table \ref{table:ft-datasets}.

\section{Prompts Used in Experiments}
\label{apdx:prompts}
The prompt used in zero-shot learning setting (Fig. \ref{fig:prompt}) and a sample prompt used in few-shot experiments with two examples (Fig. \ref{fig:few_shot_5}).

\begin{figure}
  \centering
\llmprompt{Act as an experienced security analyst, identify the TOP 3 more likely MITRE ATT\&CK Technique IDs that the text describes. Explain in 20 words why you chose these techniques specifically. Be concise. Answer in the following format: \textcolor{violet}{\{"techniques\_IDs": [YOUR\_ANSWER], "explanation": "YOUR\_EXPLANATION"\}}. 
Output only technique IDs without names. Make sure the technique IDs follow MITRE ATT\&CK Technique IDs format (i.e., TXXXX or TXXXX.YYY where X and Y are numbers from 0 to 9). Make sure the explanation matches the techniques. Make sure the output is a valid JSON.}
  \caption{\textcolor{black}{The prompt used consistently across all LLMs in the zero-shot learning setting.}}
  \label{fig:prompt}
\end{figure}

\begin{figure}
  \centering
\llmprompt{[REDACTED (see Figure 6)]

Examples:

- Input: Siloscape allows two types of instructions, one for kubectl supported commands and one for regular Windows cmd commands. | Output: \textcolor{violet}{\{"techniques\_ids": "["T1609", "T1613", "T1659"]", "explanation": "Adversaries may abuse a container administration service to execute commands within a container. A container administration service such as the Docker daemon, the Kubernetes API server, or the kubelet may allow remote management of containers within an environment."\}}

- Input: Dolphin uses a temporary scheduled task to start after installation. | Output: \textcolor{violet}{\{"techniques\_ids": "["T1053", "T1029", "T1113"]", "explanation": "Adversaries may abuse task scheduling functionality to facilitate initial or recurring execution of malicious code. Utilities exist within all major operating systems to schedule programs or scripts to be executed at a specified date and time."\}}
}
  \caption{\textcolor{black}{A sample prompt used in the few-shot learning setting with two examples.}}
  \label{fig:few_shot_5}
\end{figure}


\end{document}